\documentclass[doublecol]{epl2}
\usepackage{amssymb}
\usepackage{amsmath}
\usepackage{graphicx}%Include figure files
\usepackage{dcolumn}%Align table columns on decimal point
\usepackage{bm}
\usepackage{overpic}
\usepackage{cases}

\newcommand{\elal}{\textit{et al. }}

\title{Toward a General Understanding of the Scaling Laws in Human and Animal Mobility}
\shorttitle{Scaling Laws of Human and Animal Mobility} %Insert here a short version of the title if it exceeds 70 characters

\author{Yanqing Hu\inst{1,2} \and Jiang Zhang\inst{1} \and Di Huan\inst{1} \and Zengru Di\inst{1}}
\shortauthor{Y. Hu \etal}

\institute{
  \inst{1} Department of Systems Science, School of Management and
Center for Complexity  Research, Beijing Normal University, Beijing
100875, China\\ \inst{2} School of Mathematics, Southwest Jiaotong
University, Chengdu  610031, China}


\abstract{Recent research highlighted the scaling property of human
and animal mobility. An interesting issue is that the exponents of
scaling law for animals and human in different situations are quite
different. This paper proposes a general optimization model, a
random walker following scaling laws (whose traveling distances in
each step obey a power law distribution with exponent $\alpha$)
tries to diversify its visiting places under a given total traveling
distance with a home return probability. The results show that
different optimal exponents in between $1$ and $2$ can emerge
naturally. Therefore, the scaling property of human and animal
mobility can be understood in our framework where the discrepancy of
the scaling law exponents is due to the home return constraint under
the maximization of the visiting places diversity.}

\begin{document}

\maketitle

The scaling laws of animal mobility indicate a class of random walks
(which is also called L\'{e}vy flights), whose each step jump
distance $l$ typically follows a power-law distribution
of\begin{equation}
  \label{eq:levy}
  P(l)  \propto l^{- \alpha},
\end{equation}
where $ 1 < \alpha \leq 3 $. This kind of random walks capture the
property of the foraging behaviors\cite{viswanathan_physics_2011} of
albatrosses\cite{Edwards2007}, terrestrial
animals\cite{Viswanathan1999} and submarine
predators\cite{Sims2008}. On one hand, the exponents in animal
foraging behaviors are always approaching $2$ (e.g. $\alpha \approx
2$ for albatross\cite{Edwards2007}, deer\cite{Viswanathan1999}, and
Atlantic cod (Gadus morhua)\cite{Sims2008}, and, $\alpha \approx
1.9$ for leatherback turtle (Dermochelys coriacea), $\alpha \approx
1.7$ for Magellanic penguin (Spheniscus magellanicus)). Even for
human hunters, as Brown \elal \cite{brown_lvy_2007} pointed out, the
scaling exponent is also approaching $2$ as other foraging species.

On the other hand, as recent experimental researches highlighted,
the scaling laws (Eq.\ref{eq:levy}) can be generalized as an
approximation to the human traveling
patterns\cite{Brockmann2006,Gonzalez2008,song_modellingscaling_2010,Edwards2007}.
However, the exponents are much less than the ones of animal
foraging behaviors, such as $\alpha \approx 1.59$ in Brockmann's
bank notes tracking study\cite{Brockmann2006}, $\alpha \approx 1.2$
in Gonzalez's mobile phone users mobility\cite{Gonzalez2008} and
$\alpha \approx 1.55$ in Song's high resolution
data\cite{song_modellingscaling_2010}.

It is notable that, the difference of scaling exponents is not
negligible ($\Delta \alpha_{i}/\alpha_i \approx 25.78\%$ for
individual or $\Delta \alpha_{p}/\alpha_p \approx 66.6\%$ for
population\cite{Gonzalez2008}) and is mainly due to the behaviors in
different situations. For the human traveling or transportation
behaviors (e.g. bank notes and mobile phone users), the exponents
are approaching $1$; while for the active searching (foraging) in
the food lacking
environments\cite{bartumeus_helical_2003,humphries_environmental_2010},
the exponents are around $2$.

What is the mechanism underlying the scaling behaviors with
different exponents? On one hand, Visawanathan \elal have proposed a
model\cite{Viswanathan2000} to explain the animal's foraging
behavior, in which they assumed that the forager searches for food
``sites'' following Eq.\ref{eq:levy} with variant exponents. They
calculated the efficiency of search as the mean flights taken
between two successively visited sites and obtained the optimal
exponent $\alpha_{opt} = 2$ which yields the best searching
efficiency. Although this model can fit the foraging behaviors very
well, it fails to explain human traveling behaviors ($\alpha \approx
1$).

On the other hand, some recent studies tried to explain the human
traveling patterns, i.e., the scaling behaviors with exponent
approaching $1$. For example, Song \elal
\cite{song_modellingscaling_2010} assumed that in each time step the
individual may look for a new place which is never visited before or
go back some visited places (e.g. home or work place). This
stochastic model can explain their observed human mobility data. Han
\elal \cite{han_origin_2009} attributed the origin of human mobility
patterns to the hierarchical structure of streets. Nevertheless,
these models cannot explain the foraging behaviors.

Therefore, it still requires a general explanation. Despite the
significance of the discrepancy of scaling exponents is pointed out
by\cite{Boyer2009}, it is still poorly understood. In this paper, we
try to propose a general model of human and animal mobility.

First, we assume that one of the most important driving forces of
universal mobility patterns is diversity. That is, the random walker
``tries to'' maximize the diversity of his visiting locations (which
is measured by the Shannon entropy) within the limited total
distance\cite{jaynes_information_1957}. This constraint can be also
understood as the energy constraints of human traveling
behaviour\cite{helbing2003,buldyrev_dynamical_2003,santos_optimal_2004}.
Here, we should claim that the maximum diversity (entropy) is
achieved not by the intentional calculations of the random walker
but as a most possible result of the underlying stochastic process
\cite{jaynes_information_1957,frank_review:common_2009,banavar_applications_2010}.
Furthermore, this assumption is consistent with Visawanathan \elal
's model in the foraging behaviors because seeking foods with
highest efficiency in the food lacking environment is equivalent to
diversifying visiting places under the total distance
constraints\cite{Viswanathan2000}. And also, diversity is a very
important driving force in people's daily lives and economic
development
\cite{anderson_long_2006,page_difference:_2007,eagle_network_2010}.
This can be reflected by maximizing the diversity of visiting places
in human traveling. This assumption is also coincident with
exploring new places in Song's
model\cite{song_modellingscaling_2010}.

Second, the other important driving force exerting on the random
walker is the home return constraint. In other words, the random
walker must return home (a giving site) with a fixed probability in
our model. This assumption is also supported by our daily experience
and previous works\cite{song_modellingscaling_2010}. Both animals
and human beings always return some fixed points (homes, working
places, schools etc.).

Under the above two important presumptions, we observe that
different scaling exponents in between $1$ and $2$ can emerge
naturally under different home return probability in the process of
maximizing the diversity of visiting places. Therefore, our model
provides a possible general understanding of human and animal
mobility scaling properties.

\section{The Model}
Suppose the mobility space is an $L\times L$ toroidal lattice (where
$L$ is a given constant denoting the world size). The random walker
indicating an individual of human or animal can travel around in
this lattice according to the given scaling behavior. In other
words, the movement $\vec{\rho_t}=(x,y)$ in the $t$th time step is a
two dimensional random vector whose length $|\vec{\rho_t}|$
following Eq. \ref{eq:levy} with the exponent $\alpha$. The purpose
of this model is not to explain the origin of scaling law but to
study how the home-return tendency affects the power law exponent
$\alpha$, therefore, scaling law, i.e., Eq. \ref{eq:levy} is a basic
assumption which can be supported by the previous
studies\cite{Brockmann2006,Gonzalez2008,Viswanathan2000}.

Second, we should consider the home return constraint which is a
distinct feature of our model. In reality, human or animal
individual periodically visits or returns some specific places such
as home or working place\cite{song_modellingscaling_2010}. However,
in our model, we suppose that the random walker should return to
only one place called ``home site'' with a fixed return probability
$r$. The reasons why we select only one site as home are following:
1. To keep the model as simple as possible (The multiple fixed sites
cases will be studied in the future); 2. In the case of human
traveling, although people always travel in between home and working
place, the distance between the two points is quite small
considering the whole mobility
space\cite{song_modellingscaling_2010,Brockmann2006}. So, we can use
``home site'' to stand for the aggregation of these two points.
Without losing generality, we suppose the ``home site'' is just the
axis origin point $O=(0,0)$.

Thus, the integrate stochastic process can be described as follows.
In time step $t$, suppose the current position of the random walker
is $\vec{x_t}$. Then the next position in the time step $t+1$ is
\makeatletter
\let\@@@alph\@alph
\def\@alph#1{\ifcase#1\or \or $'$\or $''$\fi}\makeatother
\begin{subnumcases}
{\vec{x_{t+1}}=}
\vec{x_t}+\vec{\rho_t}, &if $\vec{x_t}=O$ or $\theta>r$ \label{eq:dynamics}\\
O, &if $t=0$ or $\vec{x_t}\neq O$ and $\theta\leqq r$. \label{eq:a2}
\end{subnumcases}
\makeatletter\let\@alph\@@@alph\makeatother where, $\theta$ is an
independent random number evenly distributed in $[0,1]$ and
$\vec{\rho_t}$ is an independent random vector in two dimensions
with a random length distributed following the scaling law (Eq.
\ref{eq:levy}). So the random walker starts from the origin at $t=0$
and jumps out to other places with probability $1-r$ following Eq.
\ref{eq:levy} step by step. It may return back home with a
probability $r$ in each time step.

To avoid the infinite repetition of this model, a stopping condition
is added. We assume that the total distance of this random walker
instead of the total number of time steps is given. That is, we have
the following constraint
\begin{equation}
  \label{eq:distanceconstraint}
\sum_{t=1}^T |\vec{\rho_t}| \leq W=cL,
\end{equation}
where $W$ is proportional to $L$ ($c$ is a given constant). It is
reasonable to assume that the total distance is proportional to the
total energy consumed by the random walker, so this constraint is
consistent with the energy consumptions in spatial
network\cite{helbing2003,barthelemy2011,buldyrev_dynamical_2003,santos_optimal_2004}.
Actually, for given $W$, the total time steps $T$ is a random
number. Theoretically, the distribution and expectation value of $T$
can be obtained by solving the first passage time problem, and in
this way the average stopping time $T$ is dependent on $\alpha$(see
the supplementary material). However, in the simulation, we just
generate a set of random vectors $\vec{\rho_t}$ until the total
distance constraint Eq. \ref{eq:distanceconstraint} is violated to
get $T$.

\begin{figure}[htbp]
\includegraphics[width=8.8cm]{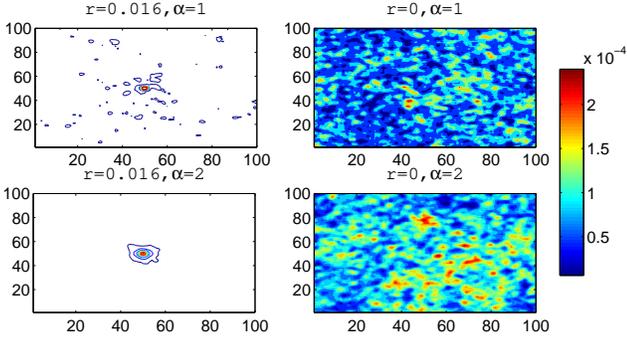}
  \caption{
  The visited probability of each site in four different cases with $L=100,W=100000$. The visited probability
distribution is comparable with the empirical results in
\cite{Gonzalez2008} qualitatively}
  \label{fig:probability}
\end{figure}

Eq. \ref{eq:levy}, \ref{eq:dynamics} and \ref{eq:distanceconstraint}
actually define the dynamical process of the random walker. Under
the dynamics, different sites may have different chances to be
visited by the random walker in his whole life. Obviously, the sites
around the home site will be visited more frequently in the tight
home return constraint (i.e., large $r$). Formally, let
$\{\vec{x_1},\vec{x_2},...,\vec{x_T}\}$ denote the sequence of sites
that the random walker have visited in all steps of one realization
of the random process. Then, the visited probability of each site
$\vec{x}$ is defined as:
\begin{equation}
  \label{eq:visitprob}
p(\vec{x})=\sum_{t=1}^T{\delta_{\vec{x_t}-\vec{x}}}/T,
\end{equation}
where, $\delta_{\vec{y}}$ is the Kronecker's delta function. It is
$1$ only when $\vec{y}=O$, otherwise it is always $0$. We can plot
visited probability of all sites in four simulations in Fig
\ref{fig:probability}. This figure can be compared to the empirical
distribution in \cite{Gonzalez2008}.

Finally, we assume that the underlying reason to form mobility
scaling patterns is the driving force of maximization of the
visiting places diversity. This trend can be quantified by
maximizing the Shannon entropy of the visited probability. The
Shannon entropy is defined as\cite{Shannon1948}:
\begin{equation}
  \label{eq:programming}
S=-\sum_{\vec{x}}^{L\times L}{p(\vec{x}) \log p(\vec{x})}.
\end{equation}
The summation is taken for all sites.

Therefore, a complete optimization problem is defined. We will find
an optimal exponent $\alpha$ to maximize Eq. \ref{eq:programming}
under the dynamical rule Eq. \ref{eq:dynamics} and total distance
constraint Eq. \ref{eq:distanceconstraint}. Obviously, different
home-return probability $r$ will systematically influence the
optimal exponent $\alpha$. Next, we will show the simulation results
of this model in the main text. And the analytic results in some
special cases are presented in the supplementary material.

\section{Results}
\begin{figure}[htbp]
\includegraphics[width=8.8cm]{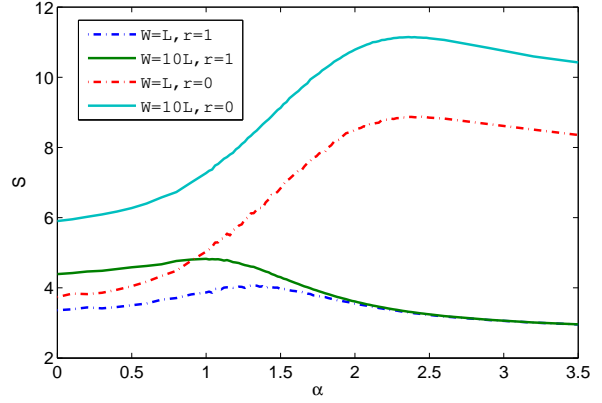}
  \caption{Entropy $S$ v.s. exponent $\alpha$ in extremal cases ($L=10000$,$W=L$ or
  $10L$). Each result on the figure is the ensemble average of
  100 experiments.  We can observe that the information entropy of visited probability will change with
  $\alpha$. The optimal $\alpha$ which can maximize $S$ appears in
  different positions with different home-return probability $r$. The
  optimal exponent is around $1$ when $r=1$ and $2$ when
  $r=0$.
  }
  \label{fig:map}
\end{figure}

Because our main purpose is to study how the home-return pattern
affects the optimal exponent. We can first consider two extremal
cases, i.e.,``never back'' case ($r=0$) and ``immediate back'' case
($r=1$). In the first case, the trajectories of the random walker
are not constraint by return probability. It means that the home
site is not effective and the random walker will travel until $W$
was fully consumed. In the second case, the random walker
immediately flies back home after one step walk. The optimal
exponents maximizing the entropy should be quite different in the
two cases. The simulation results are shown in Fig.\ref{fig:map}.

From Fig.\ref{fig:map}, we can observe that the optimal exponent
$\alpha_{ib}$ in the ``immediate back'' case ($r=1,W=L,10L$) is
around $1$. However, the optimal exponents $\alpha_{nb}$ for the
``never back'' case ($r=0,W=L,10L$) are from $2$ to $2.3$ with
different $W$. The optimal exponent in other cases of the optimized
problem should be in between of $\alpha_{ib}$ and $\alpha_{nb}$.
With simulation results in the above two extremal cases alone, we
have confirmed that the home return probability $r$ does have
influenced the scaling law.

\begin{figure}[tbp]
\includegraphics[width=8.8cm]{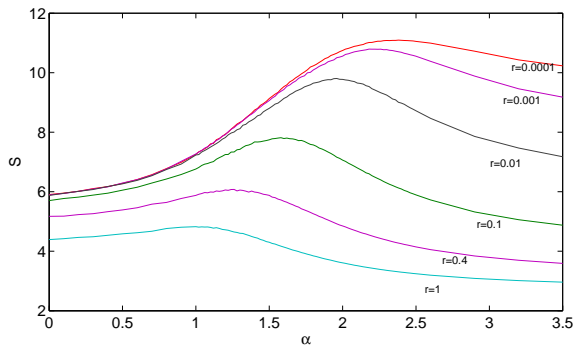}
  \caption{The dependence of entropy $S$ on $\alpha$ in different $r$($L=10000, W=10L$).
  The entropy $S$ changes with $\alpha$ and get maximal values in
  the interval around $[1,2]$ when $r$ changes from $0$ to $1$. All curves are average results of 100 experiments}
  \label{fig:entropy_alpha}
\end{figure}

\begin{figure}[tbp]
\includegraphics[width=8.8cm]{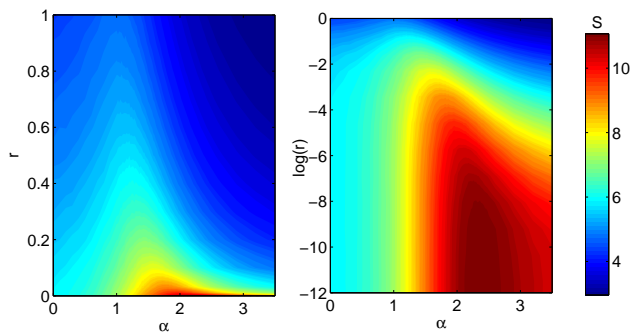}
  \caption{The dependence of entropy $S$ on $r$ and $\alpha$ ($L=10000, W=10L$).
  All curves are average results of 100 experiments}
  \label{fig:entropy}
\end{figure}

To approach how the home-return probability affects the scaling law,
we have done another group of simulations in which the overall
distance $W=10*10000$ is fixed and home return probability $r$
varies ranging from $0$ to $1$. We draw the curves showing how the
exponent $\alpha$ affects the entropy $S$ for different $r$ in
Fig.\ref{fig:entropy_alpha} and \ref{fig:entropy}.

According to Fig. \ref{fig:entropy_alpha} and \ref{fig:entropy}, the
optimal exponent $\alpha_{opt}$ changes from around $1$ to $2$ as
$r$ decreases. This result tells us that to maximize the total
entropy, a random walker seldom travel long distances when it is
more bound to the home site. And as we expected, the information
entropy $S$ decreases when $r$ increases, since random walker's
capacity of information foraging is constraint by home-return.

\begin{figure}[tbp]
\includegraphics[width=8.8cm]{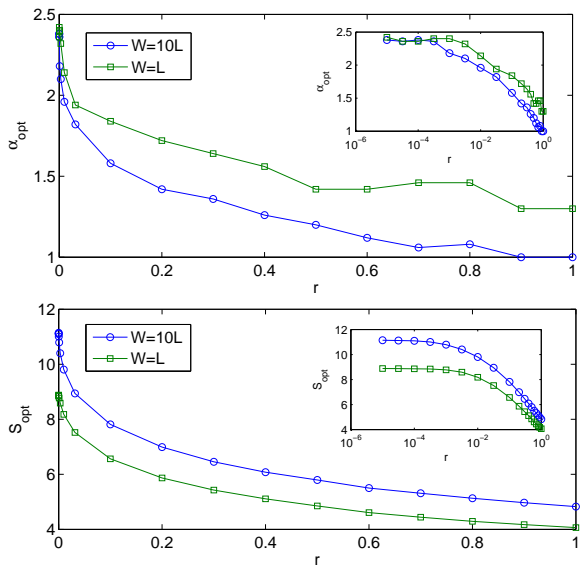}
  \caption{The dependence of optimal $\alpha$ and $S$ on $r$. Where, $L=10000$,$W=L,10L$.
  We can observe that there is an apparent transition of the optimal
  $\alpha$ from $2$ to $1.5$ when $r$ is around $[10^{-3.5},0.1]$. This
  transition is just the key factor that distinguishes human traveling and foraging behaviors. All curves are average results of 100 experiments }
  \label{fig:exponent}
\end{figure}

One may ask a question: how the home-return probability (i.e., $r$)
affects the optimal exponent $\alpha_{opt}$? To study this issue, we
conducted another group of simulations with different $r$ and plot
the curve of $(r,\alpha_{opt})$ in Fig. \ref{fig:exponent}. From the
plot we see that the value of the optimal exponent $\alpha_{opt}$
decreases slowly at first when $r$ is in $[10^{-5},10^{-4}]$ and
quickly falls down from $2$ to $1.5$ when $r$ is in
$[10^{-3.4},10^{-1}]$. At last, it drops from $1.5$ to $1$ slowly
when $r$ is larger than $0.1$. This result indicates that the
optimal scaling exponent $\alpha_{opt}$ is determined by the
home-return probability, especially when $r\in
[10^{-3.4},10^{-1}]$). It is reasonable to regard the home-return
probability $r=0.1$ as the critical point to distinguish foraging
behaviors and traveling behaviors. The interval $(0,0.1]$
corresponds to active searching behaviors and $(0.1,1]$ is for human
traveling behaviors. This critical point can be also observed in the
plot of $S_{opt}$ (the corresponding entropy in the optimal exponent
$\alpha$) versus $r$. Notice that the $S_{opt}$-$r$ curve has a
sudden change when $r<0.1$. Furthermore, from Fig.
\ref{fig:exponent} we can also observe that the optimal exponent can
be distinguished much easier in different $r$ values when $W$ is
big. So, our conclusion is much better for the asymptotic behavior
when $W$ goes infinity.

\section{Conclusion}
To summarize, we propose a general framework to understand the
mobility of human and animals based on the maximization of
information entropy and home-return constraint. Our model reveals
that the observed scaling exponents (1 for human traveling and 2 for
foraging) are the results of maximizing information entropy. Second,
our model points out that the home-return constraint is very
important and influences the optimal exponents of mobility scaling
law dramatically. The difference of scaling exponents between human
traveling behaviors and foraging behaviors is due to the strength of
home-return constraint.

As we know, human or animal mobility pattern is a key factor to
study various related topics, such as spatial networks\cite{hu2011}
gene pool diversity, mobile viruses and epidemic spreading. Although
we have not seriously fitted the simulation to empirical data, our
exploratory work contributes to mobility research by highlighting
the importance of home-return pattern based on a reasonable defined
model and large scale simulation. We inferred that the home-return
pattern is also crucial for approaching the origin of patterns for
cities, traffic systems and other related area.

% acknowledgments environment does not work as intent
\acknowledgments

We wish to thank Prof. Shlomo Havlin for some useful discussions.
The work is partially supported by NSFC under Grant No.
70771011,61004107 and 60974085.

\section{Appendix}

In the main text, we have proposed an optimal model to explain the
discrepancy of the power law exponent in human or animal mobility
patterns. However, we only presented the simulation results due to
the complexity of the model. To understand the mathematical essence
of the model and get more solid conclusions, we try to analyze this
problem in a more mathematical way.

You will see, in some extreme cases, the model can be described by
pure mathematical equations. Unfortunately, not all of these
equations are solvable, so we have to give the numeric results
instead of the exact analytic solutions. The main purpose of this
analysis is trying to give a much clearer understanding toward the
original problem but not a complete mathematical solution. We will
discuss this problem in two extreme cases according to the home
return probability $r$, namely, $r=1, r=0$.
\subsection{Immediate Return ($r=1$)}
According to our dynamical rules, when $r=1$ the random walker will
jump out following power law distribution (Eq. 1), and return back
to the ¡§home site¡¨ immediately, and then, it will jump out again
and fly back, and so on until the $T^{th}$ time step when its total
distance is consumed up.

Because each jump-out and return back cycle is independent on the
previous cycle, the final result can be converted to an equivalent
problem: $T$ random walkers start to jump one step out from the home
site simultaneously according to the scaling law (Eq. 1). Therefore,
the original temporal experiment is converted to an ensemble
experiment.

Suppose the probability of the random walker visits any site
$\vec{x}=(x,y)$ is $p(\vec{x})$, and the site $\vec{x}$ is visited
by $\eta(\vec{x})$ random walkers in the all $T$ walkers. Therefore,
the probability that $\eta(\vec{x})=i$ is a binomial distribution,
\begin{equation}
  \label{eq:binormial}
P\{\eta(\vec{x})=i\}=C_T^i{(p(\vec{x}))^i(1-p(\vec{x}))^{T-i}},
\end{equation}
Thus, the average visiting times of site $\vec{x}$ is:
\begin{equation}
  \label{eq:avgtimes}
\langle \eta(\vec{x})\rangle=Tp(\vec{x}),
\end{equation}
So, the average visit frequency of site $\vec{x}$ is:
\begin{equation}
  \label{eq:visitfrequency}
\mu(\vec{x})=\frac{\langle
\eta(\vec{x})\rangle}{\sum_{\vec{x}}{\langle
\eta(\vec{x})\rangle}}=p(\vec{x}),
\end{equation}
which is independent on the total number of walkers $T$. Then, the
Shannon entropy can be calculated as:
\begin{equation}
  \label{eq:shanentropy}
S=-\sum_{\vec{x}}^{L\times L}{p(\vec{x})\log p(\vec{x})},
\end{equation}
which is also independent on $T$. Next, we will give the concrete
mathematical form of $p(\vec{x})$ so that the relationship between
$S$ and the exponent $\alpha$ will be given. We know that
$p(\vec{x})$ is a Pareto power law distribution only when
$\alpha>1,L\rightarrow \infty$, so, we
have: \\
i.When $\alpha>1$,\\
\begin{equation}
  \label{eq:paretodistribution}
p(\vec{x})=\frac{1}{Z}|\vec{x}|^{-\alpha-1}=\frac{1}{Z}(x^2+y^2)^{-\frac{\alpha+1}{2}},
\end{equation}
for any $|\vec{x}|\geqslant 1$,where
\begin{equation}
  \label{eq:z}
  \begin{aligned}
Z&=\iint_{|\vec{x}|>1}{|\vec{x}|^{-\alpha-1}d\vec{x}}=\int_0^{2\pi}{d\theta}\int_1^{+\infty}{\rho^{-\alpha-1}\rho
d\rho}\\&=\frac{2\pi}{\alpha-1}.
\end{aligned}
\end{equation}
Notice that there are $2\pi|\vec{x}|$ points having the distance
$|\vec{x}|$ from the origin and having the visit probability
proportional to $|\vec{x}|^{-\alpha}$ , so for each point $\vec{x}$,
the visit probability should be proportional to
$|\vec{x}|^{-\alpha-1}$.

Thus the entropy can be approximated by an integration when
$L\rightarrow \infty$:
\begin{equation}
  \label{eq:s}\begin{aligned}
S&=-\iint_{|\vec{x}|>1}{p(\vec{x})\log{p(\vec{x})}d\vec{x}}\\
&=-\int_0^{2\pi}{d\theta}\int_1^L{\frac{\alpha-1}{2\pi}\rho^{-\alpha-1}\log(\frac{\alpha-1}{2\pi}\rho^{-\alpha-1})\rho
d\rho}\\
&=\frac{\alpha+1+(\alpha-1)\log \frac{2\pi}{\alpha-1}}{\alpha-1},
\end{aligned}
\end{equation}
and, we have,
\begin{equation}
  \label{eq:sdev}
\frac{\partial{S(\alpha)}}{\partial{\alpha}}=\frac{\log{\frac{2\pi}{\alpha-1}}}{\alpha-1}-\frac{\alpha+1+(\alpha-1)\log{\frac{2\pi}{\alpha-1}}}{(\alpha-1)^2},
\end{equation}
which is always smaller than 0 when $\alpha>1$. Therefore, we can
conclude that $S(\alpha)$ is a monotonic decreasing function.\\

ii. When $0<\alpha<1$\\

We know that the distribution $p(\vec{x})$ is not a standard Pareto
power law distribution but have an upper bound of $|\vec{x}|$ in
this case, so the entropy cannot be calculated as the previous case.
We will analyze the asymptotic behavior when $L\rightarrow \infty$.

At first, we know that $p(\vec{x})$ is proportional to
$|\vec{x}|^{-\alpha-1}$, therefore, Eq. \ref{eq:paretodistribution}
is still hold. However,$|\vec{x}|$ cannot go to the infinity but
have an upper bound $L$. And also, $Z$ is a normalization constant,
it is calculated as:
\begin{equation}
  \label{eq:znew}
  \begin{aligned}
Z&=\iint_{L>|\vec{x}|>1}{|\vec{x}|^{-\alpha-1}d\vec{x}}=\int_0^{2\pi}{d\theta}\int_1^{L}{\rho^{-\alpha-1}\rho
d\rho}\\&=\frac{2\pi}{\alpha-1}(1-L^{1-\alpha}).
\end{aligned}
\end{equation}
So, when $L\rightarrow \infty$, because $\alpha<1$, we know that:

\begin{equation}
  \label{eq:limitp}
  p(\vec{x})=\frac{(\alpha-1)|\vec{x}|^{-\alpha-1}}{2\pi(L^{1-\alpha}-1)}\xrightarrow[]{L\rightarrow \infty}0.
\end{equation}

Therefore, we know that the visit probability of each point
$\vec{x}$ is approaching to $0$. That implies there are no point can
be visited twice by the $T$ random walkers. In another word, we
obtain an even distribution on those $T$ visited points. So the
Shannon entropy can be calculated by the following formula:

\begin{equation}
  \label{eq:snew}
  S=-\sum_{i=1}^T{\frac{1}{T}\log{\frac{1}{T}}}=\log{T}
\end{equation}
 Instead of formula
Eq.\ref{eq:s}. In this case, we can estimate the total number of
walkers (or total time steps of one random walker) as follows.

We know for each ``jump out and return'' cycle, the average distance
traveled by the walker is:
\begin{equation}
  \label{eq:d}
  \begin{aligned}
D&=\iint_{L>|\vec{x}|>1}{|\vec{x}|\frac{1}{Z}|\vec{x}|^{-\alpha-1}d\vec{x}}\\&=\int_0^{2\pi}{d\theta}\int_1^{L}{\rho\frac{1}{Z}\rho^{-\alpha-1}\rho
d\rho}\\
&=\frac{\frac{2\pi}{\alpha-2}(1-L^{2-\alpha})}{\frac{2\pi}{\alpha-1}(1-L^{1-\alpha})}\approx
\frac{1-\alpha}{2-\alpha}L.
\end{aligned}
\end{equation}
And we know the total constraint distance can be written as,
\begin{equation}
\label{eq:distance} W=cL,
\end{equation}
where, $c$ is a constant. In the main text, we set $c=1$ or $10$ in
the simulations. So, the average total number of time steps is:
\begin{equation}
\label{eq:distance1} T=\frac{W}{D}=\frac{c(2-\alpha)}{1-\alpha},
\end{equation}
Bring it into Eq.\ref{eq:snew}, we get:
\begin{equation}
\label{eq:salpha}
S(\alpha)=\log{T}=\log{\frac{c(2-\alpha)}{1-\alpha}}.
\end{equation}
And we know:
\begin{equation}
\label{eq:partials}
\frac{\partial{S(\alpha)}}{\partial{\alpha}}=\frac{1-\alpha}{(2-\alpha)c}[\frac{(\alpha-2)c}{(\alpha-1)^2}-\frac{c}{1-\alpha}].
\end{equation}
Which is always larger than $0$ when $0<\alpha<1$. That means
$S(\alpha)$ is a monotonic increasing function. So, summarizing
cases i and ii, we know,
\begin{equation}
\label{eq:partials1}
\frac{\partial{S(\alpha)}}{\partial{\alpha}}=\left\{
\begin{aligned}&<0\ \ if\ \ \alpha>1\\&>0\ if\ \ \alpha<1
\end{aligned}\right.
\end{equation}

That implies $S(\alpha)$ can get its maximal value when $\alpha=1$.
Therefore, we have proved that when return probability $r=1$, the
visit entropy $S$ can get its maximum when $\alpha=1$ which is
consistent with our simulation result.
\subsection{Never Return ($r=0$)}
In this case, the random walker will keep jumping away from the home
site and never come back again until its energy is consumed up.

As in the case of $r=1$, assume that in each time $t$ the visit
times of a given site $\vec{x}$ is a random number $\eta(\vec{x},
t)$ which follows $0-1$ distribution, that is:
\begin{equation}
\label{eq:partials2} \eta(\vec{x},t)=\left\{
\begin{aligned}&1\ \ with\ probability\ \ p(\vec{x},t)\\&0\ \ with\ probability\ \
1-p(\vec{x},t),
\end{aligned}\right.
\end{equation}
where, $p(\vec{x},t)$ is the visit probability of the given site
$\vec{x}$ at the $t^{th}$ time step. It can be viewed as an
independent stochastic process for given $\vec{x}$. (We will get its
expression latter). Then after $T$ time steps, this site $\vec{x}$
will be visited $\xi(\vec{x})$ times,
\begin{equation}
\label{eq:xi} \xi({\vec{x}})=\sum_{t=1}^{T}{\eta(\vec{x},t)}.
\end{equation}
So, its average value is:
\begin{equation}
\label{eq:meanxi} \langle \xi({\vec{x}})\rangle
=\sum_{t=1}^{T}{\langle
\eta(\vec{x},t)\rangle}=\sum_{t=1}^{T}{p(\vec{x},t)}.
\end{equation}
Then the Shannon entropy can be calculated as:
\begin{equation}
\label{eq:bigentropy}
\begin{aligned}
S&=-\sum_{\vec{x}}^{L\times L}{\frac{\langle
\xi(\vec{x})\rangle}{\sum_{\vec{x}}{\langle
\xi(\vec{x})\rangle}}\log{\frac{\langle
\xi(\vec{x})\rangle}{\sum_{\vec{x}}{\langle
\xi(\vec{x})\rangle}}}}\\
&\begin{aligned}=&-\sum_{\vec{x}}^{L\times L}{\frac{\langle
\xi(\vec{x})\rangle}{\sum_{\vec{x}}{\langle
\xi(\vec{x})\rangle}}\log{\langle
\xi(\vec{x})\rangle}}\\&+\sum_{\vec{x}}^{L\times L}{\frac{\langle
\xi(\vec{x})\rangle}{\sum_{\vec{x}}{\langle
\xi(\vec{x})\rangle}}\log{\sum_{\vec{x}}^{L\times L}{\langle
\xi(\vec{x})\rangle}}}\end{aligned}\\
&=-\sum_{\vec{x}}^{L\times L}{\frac{\langle
\xi(\vec{x})\rangle}{\sum_{\vec{x}}{\langle
\xi(\vec{x})\rangle}}\log{\langle
\xi(\vec{x})\rangle}}+\log{\sum_{\vec{x}}^{L\times L}{\langle
\xi(\vec{x})\rangle}}\\
&\begin{aligned}=&-\sum_{\vec{x}}^{L\times
L}{\{\frac{\sum_{t=0}^{T}{p(\vec{x},t)}}{\sum_{\vec{x}}{\sum_{t=0}^{T}{p(\vec{x},t)}}}\log{\sum_{t=0}^{T}{p(\vec{x},t)}}\}}\\&+\log{\sum_{\vec{x}}^{L\times
L}{\sum_{t=0}^{T}{p(\vec{x},t)}}}\end{aligned}\\
&=-\sum_{\vec{x}}^{L\times
L}{\{\frac{\sum_{t=0}^{T}{p(\vec{x},t)}}{T}\log{\sum_{t=0}^{T}{p(\vec{x},t)}}\}}+\log{T}\\
&=-\frac{1}{T}\sum_{\vec{x}}^{L\times
L}{\{\sum_{t=0}^{T}{p(\vec{x},t)}\log{\sum_{t=0}^{T}{p(\vec{x},t)}}\}}+\log{T}
\end{aligned}
\end{equation}
Finally, we get a concise expression:
\begin{equation}
\label{eq:smu} S=-\frac{1}{T}\sum_{\vec{x}}^{L\times
L}{\mu(\vec{x},t)\log{\mu(\vec{x},t)}}+\log{T}
\end{equation}
Where,
\begin{equation}
\label{eq:mu} \mu(\vec{x},t)=\sum_{t=1}^{T}{p(\vec{x},t)}.
\end{equation}

Therefore, the Shannon entropy is just the time average value of the
entropy of the visit probability plus a constant $\log{T}$. So, $S$
is the function of $T$. However, as we know, $T$ and $p(\vec{x},t)$
are the functions of the exponent $\alpha$. Thus, we should find the
expressions $T(\alpha)$ and $\mu(\vec{x},t,\alpha)$ to solve the
problem. However, it is very hard to get the mathematical explicit
expressions. So we will only give the numeric results instead.\\

i. $T(\alpha)$\\

For given $\alpha$ and $W$ (the total distance constraint), $T$ is a
random variable. We know that in each time step, the random walker
will jump out a distance $l_t$ which is a random number with the
distance distribution $p_{l_t}(l)\propto l^{-\alpha}$, so according
to the distance constraint, we should have  the following
inequality:
\begin{equation}
\label{eq:distanceconstraint1} \sum_{t=1}^T{l_t}\leq W\ \ and\ \
\sum_{t=1}^{T+1}{l_t}> W.
\end{equation}
Then, the problem becomes a classic problem of the first passage
time of one-sided L\'{e}vy flight. We can convert this problem into
a following equivalent one: on a one-dimensional line, a random
walker starts from the origin and performs the one-sided L\'{e}vy
motion (which means the random walker can only move to right but
never back) until it will be attracted by a wall which locates at
$W$. So the time $T$ is just the first passage time to the location
$W$.

This problem has been fully discussed in the references
\cite{Koren2007,Eliazar2004}, and we can obtain the distribution of
the random variable $T$. While, here, we only need to use the
average value of $T$ as the approximation. We know
that\cite{Koren2007,Eliazar2004},

\begin{equation}
\label{eq:avgdistance} \langle T \rangle \propto W^{\alpha-1},\ \
for\ \ 1<\alpha<2.
\end{equation}

\begin{figure}[tbp]
\includegraphics[width=8.8cm]{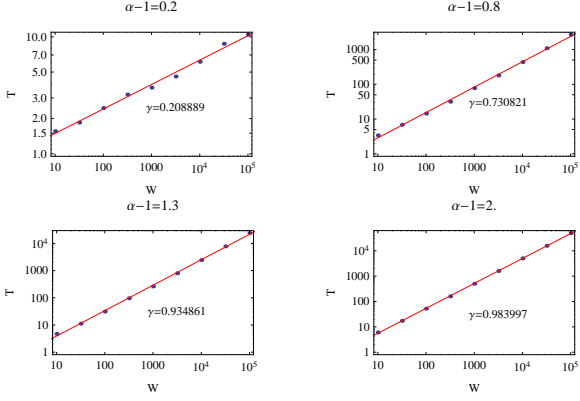}
  \caption{The Relationship between $W$ and $T$ in different $\alpha$. The slopes ($\gamma$) of the straight lines show systematic changes}
  \label{fig.gamma1}
\end{figure}

However, in our case $\alpha$ can be larger than 2. Thus, we cannot
use this analytic result. Instead, we have done a large number of
simulations about this one sided L\'{e}vy flight random walk and
found that the power law relationship between $T$ and $W$ is always
hold (see also Fig.\ref{fig.gamma1}):
\begin{equation}
\label{eq:avgtime} \langle T \rangle \propto W^{\gamma}.
\end{equation}
\begin{figure}[tbp]

While, the exponent $\gamma$ is not always $\alpha-1$ but changes
with $\alpha$ in a relationship as Fig. \ref{fig.gamma} shows.

\includegraphics[width=8.8cm]{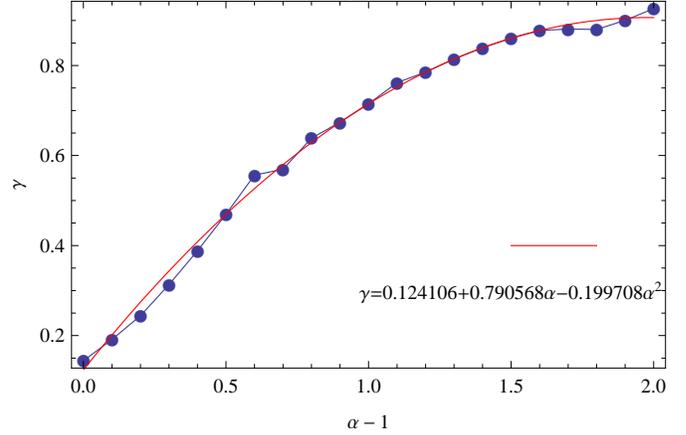}
  \caption{The Relationship between $\gamma$ and $\alpha-1$ in the one-sided L\'{e}vy flight simulations}
  \label{fig.gamma}
\end{figure}
We can use a binomial equation to approach the simulation results,
so we get the following relationship:
\begin{equation}
\label{eq:avgtime2} \begin{aligned}T(\alpha)\propto &
W^{0.124106+0.790568(\alpha-1)-0.199708(\alpha -1)^2}\\&for\ \
1<\alpha<3,
\end{aligned}
\end{equation}
Where, $T$ is just the traveling time under the distance constraint
$W$ that we will use.\\

ii.$\mu(\vec{x},t)$\\

\begin{figure}[tbp]
\includegraphics[width=8.8cm]{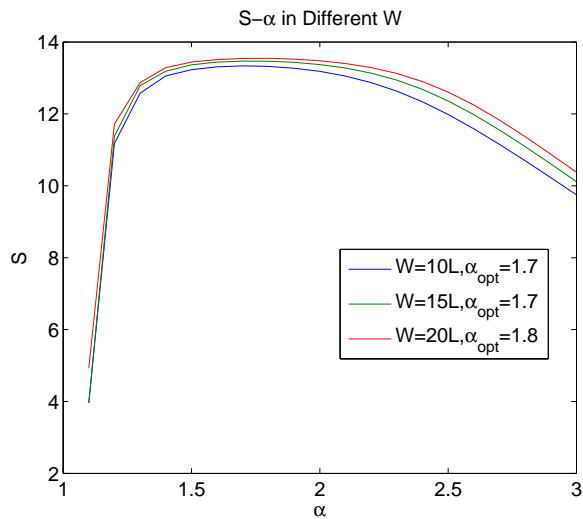}
  \caption{The relationship between $S$ and $\alpha$}
  \label{fig.final}
\end{figure}

As we know, the time continuous L\'{e}vy flight behavior can be
described by the fractional Fokker-Plank equation. That is, when $t$
is very large, $p(\vec{x},t)$ is just the approximation of the
solution of the following equation\cite{Dubkov2008}:
\begin{equation}
\label{eq:fokkerplank}
\frac{\partial{p(\vec{x},t)}}{\partial{t}}=\frac{\partial^{\alpha-1}p(\vec{x},t)}{\partial{x^{\alpha-1}}}.
\end{equation}
Where, the right hand side has the fractional differential of the
space coordinates. We can solve this equation only after the Fourier
transformation, so the solution can be written as \cite{Dubkov2008}:
\begin{equation}
\label{eq:solutionp}
p(\vec{x},t)=\frac{1}{4\pi^2}\iint{\exp{[i(\vec{k}\cdot
\vec{x})-|\vec{k}|^{\alpha-1}t]}d\vec{k}}.
\end{equation}
It is the 2-dimensional L\'{e}vy stable distribution. We know that
it has no analytic solution therefore we will only give its numeric
results. Finally,
\begin{equation}
\label{eq:solutionp1}
\mu(\vec{x},t)=\frac{1}{4\pi^2}\sum_{t=1}^{T}{\iint{\exp{[i(\vec{k}\cdot
\vec{x})-|\vec{k}|^{\alpha-1}t]}d\vec{k}}}.
\end{equation}
Then, bring this result into the Eq.\ref{eq:smu}, we can get the
numeric result of dependency of $S$ on $\alpha$ shown in Fig.
\ref{fig.final}.

In Fig.\ref{fig.final}, we set $L=1000$, $W=10,15,20L$. We can
observe that the curve can get its peak at $\alpha=1.8$  which is
close to the simulation result. Although the shape of the curve is
different from the simulation result (Fig.3) because lots of
approximations are adopted in this analysis, their main features are
similar. We can know that as the $W$ and $L$ increase, the optimal
exponent will approach to the simulation result.

\bibliography{levyflight}

\end{document}